\documentclass[11pt]{article}
\usepackage{setspace}
\usepackage{epsfig,subfigure}
\usepackage{amssymb}
\usepackage[fleqn]{amsmath}
\usepackage{arydshln}

\usepackage{color}
\newcommand{\define}{\stackrel{\triangle}{=}}
\def\QED{\mbox{\rule[0pt]{1.5ex}{1.5ex}}}

\usepackage{graphicx}
\usepackage{color}

\usepackage{amssymb}
\usepackage{amsmath,amsfonts,amssymb}
\usepackage{verbatim}
\usepackage{stfloats}
\usepackage[bookmarks=false]{}

\newtheorem{theorem}{Theorem}

\newtheorem{lemma}{Lemma}

\newcommand\blfootnote[1]{%
  \begingroup
  \renewcommand\thefootnote{}\footnote{#1}%
  \addtocounter{footnote}{-1}%
  \endgroup
}

\begin{document}
\date{}
\title{%Blind Interference Alignment for \\ Private Information Retrieval
The Capacity of Robust Private Information Retrieval \\
with Colluding Databases
%\thanks{This work is supported by NSF grants CCF-1317351 and CCF-0963925.}
}
%\author{\IEEEauthorblockN{Hua Sun and Syed A. Jafar}
%\IEEEauthorblockA{Center for Pervasive Communications and Computing (CPCC)\\
%University of California Irvine, Irvine, CA 92697
%%\\ Email: \{huas2, syed\}@uci.edu
%}}
\author{ \normalsize Hua Sun and Syed A. Jafar \\
%{\small Center for Pervasive Communications and Computing (CPCC)}\\
%{\small University of California Irvine, Irvine, CA 92697}\\
%{\small \it Email: \{huas2, syed\}@uci.edu}
}

\maketitle

\blfootnote{Hua Sun (email: huas2@uci.edu) and Syed A. Jafar (email: syed@uci.edu) are with the Center of Pervasive Communications and Computing (CPCC) in the Department of Electrical Engineering and Computer Science (EECS) at the University of California Irvine. %The work is supported by grants from  ONR, NSF and ARL. The results of this work were submitted in part for presentation at IEEE ISIT 2016 and IEEE GLOBECOM 2016. 
}

\begin{abstract}
Private information retrieval (PIR)  is the problem of retrieving as efficiently as possible, one out of $K$ messages from $N$ non-communicating replicated databases (each holds all $K$ messages) while keeping the identity of the desired message index a secret from each individual database. The information theoretic capacity of  PIR (equivalently, the reciprocal of minimum download cost) is the maximum number of bits of desired information that can be privately retrieved per bit of downloaded information. $T$-private PIR is a generalization of PIR to include the requirement that even if any $T$ of the $N$ databases collude, the identity of the retrieved message remains completely unknown to them. Robust PIR is another generalization that refers to the scenario where we have $M \geq N$ databases, out of which any $M - N$  may fail to respond.  For $K$ messages and $M\geq N$ databases out of which at least some $N$ must respond, we show that the capacity of $T$-private and Robust PIR  is $\left(1+T/N+T^2/N^2+\cdots+T^{K-1}/N^{K-1}\right)^{-1}$. The  result includes as special cases the capacity of PIR without robustness  ($M=N$) or $T$-privacy constraints ($T=1$).
\end{abstract}

%\newpage
\allowdisplaybreaks
\section{Introduction}
The private information retrieval (PIR) problem is motivated by the desire to protect the privacy of a user against  data providers. Besides its direct applications in data privacy, it is intimately related to many fundamental problems in cryptography, e.g., oblivious transfer \cite{SymPIR}, instance hiding \cite{Hide, Hide_one, Hide_multiple},  secure multiparty computation \cite{Local_random}, and secret sharing schemes \cite{Shamir, Beimel_Ishai_Kushilevitz_Orlov}. The significance of PIR also extends  beyond security, through its fundamental connections to other prominent topics such as locally decodable codes \cite{YekhaninPhd} and batch codes \cite{batch} in coding theory, relationships between communication and computation  \cite{Ishai_Kushilevitz} in complexity theory, and most recently blind interference alignment \cite{Sun_Jafar_BIAPIR} in wireless communications. In fact most constructions of locally decodable codes are translated directly from PIR schemes. Through the connections between locally decodable and locally recoverable codes \cite{Gopalan_Huang_Simitci_Yekhanin}, PIR also connects to  distributed data storage repair \cite{Dimakis_survey} and index coding \cite{Birk_Kol_Trans}, which in turn encompass all of  network coding \cite{Ahlswede_Cai_etal}. Therefore PIR represents an important focal point to tackle significant challenges across these fields.

The goal of PIR is to find the most efficient way for a user to retrieve a desired message from a set of $N$ distributed databases, each of which stores all $K$ messages, without revealing anything (in the information theoretic sense\footnote{There is another line of research, where privacy needs to be satisfied only for computationally bounded databases \cite{William, Yekhanin, CPIR}.}) about which message is being retrieved,  to any individual database. The PIR problem was initially studied in the setting where each message is one bit long \cite{PIRfirst, PIRfirstjournal, Ambainis, Beimel_Ishai_Kushilevitz_Raymond, YekhaninPhd, 2PIR}, where the cost of a PIR scheme is  measured by the total amount of communication between the user and the databases, i.e., the sum of communications from the user to the databases (upload) and from the databases to the user (download). What is pursued in this work is the traditional Shannon theoretic formulation, where message size is allowed to be arbitrarily large, and therefore the upload cost is negligible compared to the download cost \cite{Chan_Ho_Yamamoto, PIRfirstjournal}. The information theoretic capacity  of PIR is the maximum number of bits of desired information that can be privately retrieved per bit of downloaded information. Equivalently, it is the reciprocal of the minimum possible download cost per bit of desired message.  In \cite{Sun_Jafar_PIR}, we showed that the information theoretic capacity of PIR, for arbitrary number of messages $K$ and {arbitrary} number of databases $N$ is $\left(1+1/N+1/N^2+\cdots+1/N^{K-1}\right)^{-1}$. 

There are several interesting extensions of PIR that explore its limitations under additional constraints. These include extensions where up to $T$ of the $N$ databases may collude \cite{Beimel_Ishai_Kushilevitz, Barkol_Ishai_Weinreb} ($T$-private PIR); where some of the databases may not respond \cite{Beimel_Stahl} (Robust PIR); where both the privacy of the user and the databases must be protected \cite{SymPIR} (Symmetric PIR); where only one database holds all the messages and all other databases hold independent information \cite{Gertner_Goldwasser_Malkin}; where retrieval operations are unsynchronized \cite{Fanti_Ramchandran}; and where beyond communications, computation is also a concern \cite{Beimel_Ishai_Malkin}. There is also much recent work in the distributed storage setting \cite{Shah_Rashmi_Kannan, Chan_Ho_Yamamoto, Fazeli_Vardy_Yaakobi, Tajeddine_Rouayheb} (the databases form a distributed storage system) where the main focus is on how  the coding of the storage system works jointly with PIR. 

In this work, we mainly consider $T$-private PIR in the Shannon theoretic setting, where we have an arbitrary number of messages ($K$), arbitrary number of databases ($N$), each database stores all the messages, the messages are allowed to be arbitrarily large, and the privacy of the desired message index must be guaranteed even if any $T$ of the $N$ databases collude. The main contribution of this work is to show that the information theoretic capacity of  $T$-private PIR is $\left(1+T/N+T^2/N^2+\cdots+T^{K-1}/N^{K-1}\right)^{-1}$.

We further consider the extension to \emph{robust} $T$-private PIR, where we have $M \geq N$ databases, out of which any $M - N$ databases may not respond, so that with answers from any $N$ databases, we need to ensure both privacy and correctness. In this context, the contribution of this work is to show that the information theoretic capacity of robust $T$-private PIR remains the same as that of $T$-private PIR, i.e., there is no capacity cost from not knowing in advance  \emph{which} $N$ databases will respond.

{\it Notation: For  $n_1, n_2\in\mathbb{Z}, n_1\leq n_2$, define the notation $[n_1:n_2]$ as the set $\{n_1,n_1+1,\cdots, n_2\}$ and $(n_1:n_2)$ as the vector $(n_1, n_1+1, \cdots, n_2)$.  For an index set $\mathcal{I}=\{i_1, i_2, \cdots, i_n\}$, the notation $A_{\mathcal{I}}$ represents the set $\{A_i:  i\in\mathcal{I}\}$. For an index vector $\mathcal{I}=(i_1, i_2, \cdots, i_n)$, the notation $A_\mathcal{I}$ represents the vector $(A_{i_1}, A_{i_2}, \cdots, A_{i_n})$. For a matrix $S$, the notation $S[\mathcal{I}, :]$ represents the submatrix of $S$ formed by retaining only the rows corresponding to the elements of the vector $\mathcal{I}$. The notation $X\sim Y$ is used to indicate that $X$ and $Y$ are identically distributed.}

\section{Problem Statement}\label{sec:model}
\subsection{$T$-private PIR}
Consider $K$ independent messages $W_1, \cdots, W_K$ of size $L$ bits each. 
\begin{eqnarray}
&& H(W_1, \cdots, W_K) = H(W_1) + \cdots + H(W_K), \label{h1}\\
&& H(W_1) = \cdots = H(W_K) = L. \label{h2}
\end{eqnarray}
There are $N$ databases. Each database stores all the messages $W_1, \cdots, W_K$. A user wants to retrieve $W_k, k \in [1:K]$ subject to $T$-privacy, i.e., without revealing anything about the message identity, $k$, to any colluding susbset of up to $T$ out of the $N$ databases.

To retrieve $W_k$ privately, the user  generates $N$ queries $Q_1^{[k]}, \cdots, Q_N^{[k]}$, where the superscript denotes the desired message index. Since the queries are generated with no knowledge of the realizations of the messages, the queries must be independent of the messages,
\begin{eqnarray}
I(W_1, \cdots, W_K; Q_1^{[k]}, \cdots, Q_N^{[k]}) = 0 \label{qwind}.
\end{eqnarray}
The user  sends query $Q_n^{[k]}$ to the $n$-th database, $\forall n\in[1:N]$. Upon receiving $Q_n^{[k]}$, the $n$-th database generates an answering string $A_n^{[k]}$, which is a deterministic function of $Q_n^{[k]}$ and the data stored (i.e., all messages $W_1, \cdots, W_K$), 
\begin{eqnarray}
H(A_n^{[k]} | Q_n^{[k]}, W_1, \cdots, W_K) = 0. \label{ansdet}
\end{eqnarray}
Each database returns to the user its answer $A_n^{[k]}$. From all answers $A_1^{[k]}, \cdots, A_N^{[k]}$, the user can decode the desired message $W_k$,%, with probability of error $P_e$. 
%In particular, we require that the probability of error approaches zero when the size of each message, $L$ approaches infinity. From Fano's inequality, we have
\begin{eqnarray}
\mbox{[Correctness]} ~H(W_k | A_1^{[k]}, \cdots, A_N^{[k]}, {\color{black} Q_1^{[k]}, \cdots, Q_N^{[k]}}) = 0.%L \epsilon_L \label{corr}
\end{eqnarray}
%where $\epsilon_L$ approaches zero when $L$ approaches infinity.

To satisfy the privacy constraint that any $T$ colluding databases learn nothing about the desired message index $k$ information theoretically, any $T$ queries must be independent of $k$. Let $\mathcal{T}$ be a subset of $[1:N]$ and its cardinality be denoted by $|\mathcal{T}|$. $Q_{\mathcal{T}}^{[k]}$ represents the subset $\{Q_n^{[k]}, n \in \mathcal{T}\}$. $A_{\mathcal{T}}^{[k]}$ is defined similarly. To satisfy the $T$-privacy requirement we must have %$I(Q_j^{[i]}; i) = 0, \forall j.$
\begin{eqnarray}
\mbox{[Privacy]} ~~~ I(Q_{\mathcal{T}}^{[k]}; k) = 0, \forall \mathcal{T} \subset [1:N], |\mathcal{T}| = T. \label{qi}
\end{eqnarray}

As the answering string is a deterministic function of the query and all messages, any set of $T$ answering strings must be independent of $k$ as well, 
\begin{eqnarray}
I(A_{\mathcal{T}}^{[k]}; k) = 0, \forall \mathcal{T} \subset [1:N], |\mathcal{T}| = T. \label{ai}
\end{eqnarray}
To underscore that any set of $T$ or fewer answering strings is independent of the desired message index, we may suppress the superscript and write $A_{\mathcal{T}}$ directly instead of $A_{\mathcal{T}}^{[k]}$, and express the elements of such a set as $A_n$ instead of $A_n^{[k]}$.

The metric that we study in this paper is the PIR rate, which characterizes how many bits of desired information are retrieved per downloaded bit. Note that the PIR rate is the reciprocal of download cost.  The rate $R$ of a PIR scheme is defined as follows.
\begin{eqnarray}
R \define 
%\lim_{L \rightarrow \infty} 
%\frac{H(W_i)}{\sum_{n=1}^N H(A_n^{[i]})} = 
\frac{L}{\sum_{n=1}^N H(A_n)} . \label{eta_def}
\end{eqnarray}
The capacity, $C$, is the supremum of $R$ over all PIR schemes.
%A rate $R$ is said to be zero-error achievable if there exists a PIR scheme of rate greater than or equal to $R$ for which $P_e=0$. It is said to be $\epsilon$-error achievable if there exists a sequence of PIR schemes, each of rate greater than or equal to $R$, for which $P_e\rightarrow 0$ as $L\rightarrow\infty$. The supremum of zero-error achievable rates is called the zero-error capacity, $C_o$, and the supremum of $\epsilon$-error achievable rates is called the $\epsilon$-error capacity $C_\epsilon$. Our goal is to characterize both the zero-error capacity $C_o$ and the $\epsilon$-error capacity, $C_\epsilon$, of $T$-private PIR. 

\subsection{Robust $T$-private PIR}
The robust $T$-private PIR problem is defined similar to the $T$-private PIR problem. The only difference is that instead of $N$ databases, we have $M \geq N$ databases, and the correctness condition needs to be satisfied when the user collects \emph{any} $N$ out of the $M$ answering strings. 
\section{Main Result: Capacity of Robust $T$-Private PIR
}\label{sec:main}
The following theorem states the main result.
\begin{theorem}\label{thm:download}
For  $T$-private PIR with $K$ messages and $N$ databases, the capacity is
\begin{eqnarray}
C = \left(1 +T/N + T^2/{N^2} + \cdots +T^{K-1}/{N^{K-1}}\right)^{-1}.
\end{eqnarray}
\end{theorem}

The capacity of PIR with $T$ colluding databases generalizes the case without $T$-privacy constraints, where $T = 1$ \cite{Sun_Jafar_PIR}. The capacity is a strictly decreasing function of $T$. When $T = N$, the capacity is $1/K$, meaning that the user has to download all $K$ messages to be private, as in this case, the colluding databases are as strong as the user. Similar to the $T = 1$ case, the capacity is strictly deceasing in the number of messages, $K$, and strictly increasing in the number of databases, $N$. When the number of messages approaches infinity, the capacity approaches $1 - T/N$, and when the number of databases approaches infinity ($T$ remains constant), the capacity approaches 1. Finally, note that since the download cost is the reciprocal of the rate, the capacity characterization in Theorem \ref{thm:download}  equivalently  characterizes the optimal download cost per message bit for $T$-private PIR as $\left(1 + T/N + T^2/{N^2} + \cdots + T^{K-1}/{N^{K-1}}\right)$ bits.

The capacity-achieving scheme that we construct for  $T$-private PIR, generalizes easily to incorporate robustness constraints. As a consequence, we are also able to characterize the capacity of robust $T$-private PIR. This result is stated in the following theorem. 
\begin{theorem}\label{thm:robust}
The capacity of robust $T$-private PIR is \begin{eqnarray}
C = \left(1 +T/N + T^2/{N^2} + \cdots +T^{K-1}/{N^{K-1}}\right)^{-1}.
\end{eqnarray}
\end{theorem}
Since the capacity expressions are the same, we note that there is no capacity penalty from not knowing in advance which $N$ databases will respond. Even though this uncertainty increases as $M$ increases, capacity is not a function of $M$. However, we note that the communication complexity of our capacity achieving scheme does increase with $M$.

%The proof for the general case of $K$ messages, and $N$ databases is presented in Section \ref{sec:proof}. In order to convey the main ideas, we begin here with some simple examples. 

\section{Proof of Theorem \ref{thm:download}: Achievability}\label{sec:ach}
There are two key aspects of the achievable scheme -- 1) the query structure, and 2) the specialization of the query structure to ensure $T$-privacy and correctness. While the query structure is  different from the $T=1$ setting of \cite{Sun_Jafar_PIR}, it draws upon the iterative application of the same three principles that were identified in \cite{Sun_Jafar_PIR}. These principles are listed below.
\begin{enumerate}
\item[(1)] {\it Enforcing Symmetry Across Databases}
\item[(2)] {\it Enforcing Message Symmetry within the Query to Each Database}
\item[(3)] {\it Exploiting  Previously Acquired Side Information of Undesired Messages to Retrieve New Desired Information}
\end{enumerate}
The specialization of the structure to ensure $T$-privacy and correctness is another novel element of the achievable scheme. To illustrate how these ideas work together in an iterative fashion, we will present a few simple examples corresponding to small values of $K, N$ and $T$, and then generalize it to arbitrary $K, N$ and $T$. Let us begin with a lemma.

\begin{lemma}\label{lemma:inv}
Let $S_1, S_2, \cdots, S_K \in \mathcal{F}_q^{\alpha \times \alpha}$ be $K$ random matrices, drawn independently and uniformly from all $\alpha \times \alpha$ full-rank matrices over $\mathcal{F}_q$. Let $G_1, G_2, \cdots, G_K \in \mathcal{F}_q^{\beta \times \beta}$ be  $K$ invertible square matrices of dimension $\beta\times \beta$ over $\mathcal{F}_q$. Let $\mathcal{I}_1, \mathcal{I}_2, \cdots, \mathcal{I}_K \in \mathbb{N}^{\beta \times 1}$ be $K$ index vectors, each containing $\beta$ distinct indices from $[1:\alpha]$. Then 
\begin{eqnarray}
(G_1 S_1[\mathcal{I}_1,:],G_2 S_2[\mathcal{I}_2,:], \cdots,G_K S_K[\mathcal{I}_K,:] ) \sim  (S_1[(1:\beta),:],S_2[(1:\beta),:], \cdots, S_K[(1:\beta),:] )\label{eq:id}
\end{eqnarray}
where $S_i[\mathcal{I}_i,:], i \in [1:K]$ are $\beta \times \alpha$ matrices comprised of the rows of $S_i$ with indices in $\mathcal{I}_i$.
\end{lemma}

{\it Proof:} We wish to prove that the left hand side of (\ref{eq:id}) is identically distributed (recall that the notation $X\sim Y$ means that $X$ and $Y$ are identically distributed) to the right hand side of (\ref{eq:id}). 
Because the rank of a matrix does not depend on the ordering of the rows, we have $$(S_1[\mathcal{I}_1,:],S_2[\mathcal{I}_2,:], \cdots, S_K[\mathcal{I}_K,:] ) \sim  (S_1[(1:\beta),:],S_2[(1:\beta),:], \cdots, S_K[(1:\beta),:] )$$ Since $S_i$ are picked uniformly from all full-rank matrices, conditioned on any feasible value of the remaining rows $S_i[(\beta+1:\alpha),:]$, the first $\beta$ rows $S_i[(1:\beta),:]$  are uniformly distributed over all possibilities that preserve full-rank for $S_i$. Now note that the mapping from $S_i[(1:\beta),:]$ to $G_iS_i[(1:\beta),:]$ is bijective, and $S_i[(1:\beta),:]$ spans the same row space as $G_iS_i[(1:\beta),:]$, i.e., replacing $S_i[(1:\beta),:]$ with $G_iS_i[(1:\beta),:]$, preserves $S_i$  as a full-rank matrix.  Therefore, conditioned on any feasible $S_i[(\beta+1:\alpha),:]$, the set of feasible values of $S_i[(1:\beta),:]$  is the same as the set of feasible $G_iS_i[(1:\beta),:]$ values. Therefore, $G_iS_i[(1:\beta),:]$ is also uniformly distributed over the same set. Finally,  since the $S_i$ are chosen independently, the statement of Lemma \ref{lemma:inv} follows.
\hfill\QED

\subsection{$K = 2$ Messages, $N = 3$ Databases, $T = 2$ Colluding Databases}\label{sec:k2n3t2}
The capacity for this setting, is $C=\left(1+\frac{2}{3}\right)^{-1}=\frac{3}{5}$.
\subsubsection{Query Structure}
We begin by constructing a query structure, which will then be specialized to achieve correctness and privacy. Without loss of generality, let $[a_k]$ denote the symbols of the desired message, and $[b_k]$ the symbols of the undesired message.
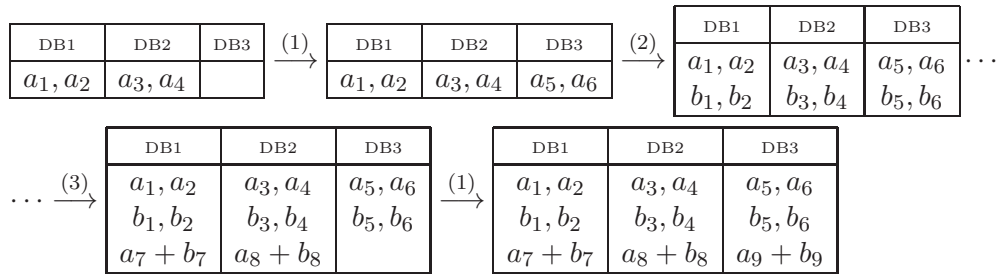
\begin{eqnarray*}
&&\begin{array}{|c|c|c|c|c|}\hline
\mbox{\tiny DB1}&\mbox{\tiny DB2}&\mbox{\tiny DB3}\\\hline
a_1,a_2&a_3,a_4&~~\\\hline
\end{array}\stackrel{(1)}{\longrightarrow}\begin{array}{|c|c|c|c|c|}\hline
\mbox{\tiny DB1}&\mbox{\tiny DB2}&\mbox{\tiny DB3}\\\hline
a_1,a_2&a_3,a_4&a_5,a_6\\\hline
\end{array}\stackrel{(2)}{\longrightarrow}\begin{array}{|c|c|c|c|c|}\hline
\mbox{\tiny DB1}&\mbox{\tiny DB2}&\mbox{\tiny DB3}\\\hline
a_1,a_2&a_3,a_4&a_5,a_6\\
b_1,b_2&b_3,b_4&b_5,b_6\\\hline
\end{array}\cdots\\
&&\cdots\stackrel{(3)}{\longrightarrow}\begin{array}{|c|c|c|c|c|}\hline
\mbox{\tiny DB1}&\mbox{\tiny DB2}&\mbox{\tiny DB3}\\\hline
a_1,a_2&a_3,a_4&a_5,a_6\\
b_1,b_2&b_3,b_4&b_5,b_6\\
a_7+b_7&a_8+b_8&\\
\hline
\end{array}
\stackrel{(1)}{\longrightarrow}\begin{array}{|c|c|c|c|c|}\hline
\mbox{\tiny DB1}&\mbox{\tiny DB2}&\mbox{\tiny DB3}\\\hline
a_1,a_2&a_3,a_4&a_5,a_6\\
b_1,b_2&b_3,b_4&b_5,b_6\\
a_7+b_7&a_8+b_8&a_9+b_9\\
\hline
\end{array}
\end{eqnarray*}
We start  by requesting the first $T^{K-1}=2$ symbols from each of the first $T=2$ databases: $a_1, a_2$ from DB1, and $a_3,a_4$ from DB2. Applying database symmetry, we  simultaneously request  $a_5, a_6$ from DB3. Next, we enforce message symmetry, by including queries for $b_1, \cdots, b_6$ as the counterparts for $a_1,\cdots, a_6$. 
Now consider the first $T=2$ databases, i.e., DB1 and DB2, which can potentially collude with each other. Unknown to these databases the user has acquired two symbols of external side information, $b_5, b_6$, comprised of undesired message symbols received from DB3. Splitting the two symbols of external side information among DB1 and DB2 allows the user one symbol of side information for each of DB1 and DB2 that it can exploit to retrieve new desired information symbols. In our construction of the query structure, we will assign new labels to the external side-information exploited within each database, e.g., $b_7$ for DB1 and $b_8$ for DB2, with the understanding that eventually when the dependencies within the structure are specialized, $b_7, b_8$ will turn out to be  functions of previously acquired side-information. Using its assigned side information, each DB  acquires a new symbol of desired message, so that DB1 requests $a_7+b_7$ and DB2 requests $a_8+b_8$. Finally, enforcing symmetry across databases,  DB3 requests $a_9+b_9$. At this point, the construction is symmetric across databases, the query to any database is symmetric in itself across messages, and the amount of side information exploited within any $T$ colluding databases equals the amount of side information available external to those $T$ databases. So the skeleton of the query structure is complete.

Note that if DB1 and DB2 collude, then the external side information is $b_5, b_6$, so we would like the side-information that is exploited by DB1 and DB2, i.e., $b_7, b_8$ to be functions of the external side information that is available, i.e., $b_5, b_6$. However, since \emph{any} $T=2$ databases can collude, it is also possible that DB1 and DB3 collude instead, in which case we would like $b_7, b_9$ to be functions of side information that is external to DB1 and DB3, i.e., $b_3,b_4$. Similarly, if DB2 and DB3 collude, then we would like $b_8, b_9$ to be functions of $b_1,b_2$. How to achieve such dependencies in a manner that preserves privacy and ensures correctness is the remaining challenge. Intuitively, the key is to make $b_7, b_8, b_9$ depend on \emph{all} side-information $b_1,b_2, \cdots, b_6$ in a generic sense. In other words, we will achieve the desired functional dependencies by viewing $b_1, b_2, \cdots, b_9$ as the outputs of a $(9,6)$ MDS code, so that any $3$ of these $b_k$ are functions of the remaining $6$. The details of this specialization are described next.

\subsubsection{Specialization to Ensure Correctness and Privacy}
Let each message consist of $N^K=9$ symbols from a sufficiently large\footnote{The requirements on the size of the field have to do with the existence of MDS codes that are used in the construction. In this case $q\geq N^K$ is sufficient.} finite field $\mathbb{F}_q$. The messages  $W_1$, $W_2\in\mathbb{F}_q^{9\times 1}$ are then represented as $9\times 1$ vectors over $\mathbb{F}_q$. Let $S_1, S_2\in\mathbb{F}_q^{9\times 9}$ represent random matrices chosen privately by the user, independently and uniformly from all $9\times 9$  full-rank matrices over $\mathbb{F}_q$. Without loss of generality, let us assume that $W_1$ is the desired message. Define the $9\times 1$ vectors $a_{[1:9]}\in \mathbb{F}_q^{9\times 1}$ and $b_{[1:9]}\in \mathbb{F}_q^{9 \times 1}$, as follows
\begin{eqnarray}
a_{[1:9]}&=&S_1W_1\\
b_{[1:9]}&=&\mbox{MDS}_{9\times 6}S_2[(1:6),:]W_2
\end{eqnarray}
where $S_2[(1:6),:]$ is a $6\times 9$ matrix comprised of the first $6$ rows of $S_2$. MDS$_{9\times 6}$ is the generator matrix of a $(9, 6)$ MDS code (e.g., a Reed Solomon code). The generator matrix does not need to be random, i.e., it may be globally known. Note that because of the MDS property, any $6$ rows of MDS$_{9\times 6}$ form a $6\times 6$ invertible matrix. Therefore, from any $6$ elements of $b_{[1:9]}$, all $9$ elements of $b_{[1:9]}$ can be recovered. For example, from $b_1, b_2, \cdots, b_6$, one can recover $b_7, b_8, b_9$. The queries from each database are constructed according to the structure described earlier.

\begin{eqnarray}
\begin{array}{|c|c|c|c|c|}\hline
\mbox{\tiny DB1}&\mbox{\tiny DB2}&\mbox{\tiny DB3}\\\hline
a_1,a_2&a_3,a_4&a_5,a_6\\
b_1,b_2&b_3,b_4&b_5,b_6\\
a_7+b_7&a_8+b_8&a_9+b_9\\
\hline
\end{array}
\end{eqnarray}
Correctness is easy to see, because the user recovers $b_{[1:6]}$ explicitly, from which it can recover all $b_{[1:9]}$, thereby allowing it to recover all of $a_{[1:9]}$. Let us see why privacy holds. The queries for any $T=2$ colluding databases are comprised of $6$ variables from $a_{[1:9]}$ and $6$ variables from $b_{[1:9]}$. Let the indices of these variables be denoted by the $6\times 1$ vectors $\mathcal{I}_a,\mathcal{I}_b\in\mathbb{N}^{6\times 1}$, respectively, so that,

\begin{eqnarray}
(a_{\mathcal{I}_a}, b_{\mathcal{I}_b})&=&(S_1[\mathcal{I}_a,:]W_1,\mbox{MDS}_{9\times 6}[\mathcal{I}_b,:]S_2[(1:6),:]W_2)\\
&\sim&(S_1[(1:6),:]W_1,S_2[(1:6),:]W_2)\label{eq:sim}
\end{eqnarray}
where (\ref{eq:sim}) follows from Lemma \ref{lemma:inv} because $\mbox{MDS}_{9\times 6}[\mathcal{I}_b,:]$ is an invertible $6\times 6$ matrix. Therefore, the random map from $W_1$ to $a_{\mathcal{I}_a}$ variables is i.i.d. as the random map from $W_2$ to $b_{\mathcal{I}_b}$, and privacy is guaranteed.
Note that since $9$ desired symbols are recovered from a total of $15$ downloaded symbols, the rate achieved by this scheme is $9/15=3/5$, which matches the capacity for this setting. While this specialization suffices for our purpose (it achieves capacity), we  note that further simplifications of the scheme are possible, which allow it to operate over smaller fields and with lower upload cost. Such an example is provided in the conclusion section of this paper.

%Note that $\mbox{MDS}_{9\times 6}[\mathcal{I}_b,:]$ is an invertible $6\times 6$ matrix, so the mapping from $S_2[(1:6),:]$ to $\mbox{MDS}_{9\times 6}[\mathcal{I}_b,:]S_2[(1:6),:]$ is bijective. Also, $S_2[(1:6),:]$ spans the same row space as $\mbox{MDS}_{9\times 6}[\mathcal{I}_b,:]S_2[(1:6),:]$. Therefore, if $S_2$ is full rank, then replacing its first 6 rows, i.e., $S_2[(1:6),:]$ with $\mbox{MDS}[\mathcal{I}_b,:]S_2[(1:6),:]$, preserves it as a full-rank matrix. Therefore, $S_2[(1:6),:]$ is identically distributed as $\mbox{MDS}_{9\times 6}[\mathcal{I}_b,:]S_2[(1:6),:]$. Further, since the rank of a matrix is not dependent on the ordering of rows, $\mbox{MDS}_{9\times 6}[\mathcal{I}_b,:]S_2[(1:6),:]$ is independent and identically distributed to $S_1[\mathcal{I}_a,:]$. Therefore, the random map from $W_1$ to $a_{\mathcal{I}_a}$ variables is i.i.d. as the random map from $W_2$ to $b_{\mathcal{I}_b}$, and privacy is guaranteed.

\subsection{$K = 2$ Messages, $N = 4$ Databases, $T = 2$ Colluding Databases}
The capacity for this setting, is $C=\left(1+\frac{2}{4}\right)^{-1}=\frac{2}{3}$.
\subsubsection{Query Structure}
The query structure is constructed as follows.
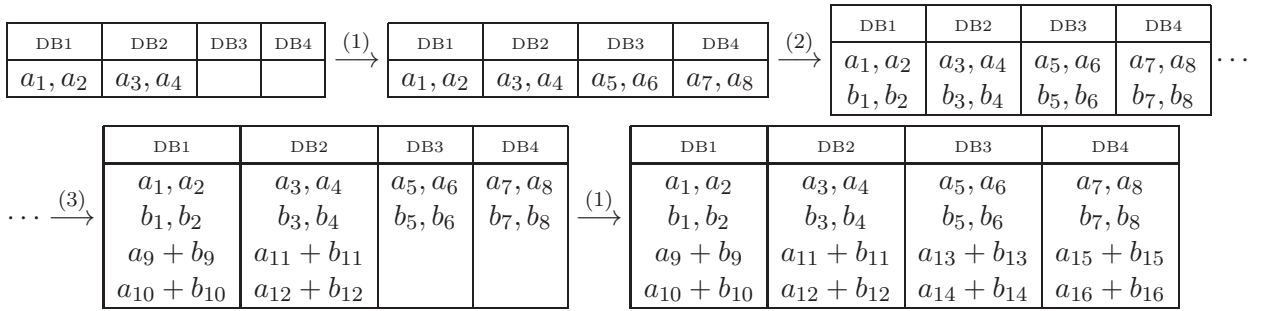
\begin{eqnarray*}
&&\begin{array}{|c|c|c|c|c|c|}\hline
\mbox{\tiny DB1}&\mbox{\tiny DB2}&\mbox{\tiny DB3}&\mbox{\tiny DB4}\\\hline
a_1,a_2&a_3,a_4&~~&~~\\\hline
\end{array}\stackrel{(1)}{\longrightarrow}\begin{array}{|c|c|c|c|c|}\hline
\mbox{\tiny DB1}&\mbox{\tiny DB2}&\mbox{\tiny DB3}&\mbox{\tiny DB4}\\\hline
a_1,a_2&a_3,a_4&a_5,a_6&a_7,a_8\\\hline
\end{array}\stackrel{(2)}{\longrightarrow}\begin{array}{|c|c|c|c|c|}\hline
\mbox{\tiny DB1}&\mbox{\tiny DB2}&\mbox{\tiny DB3}&\mbox{\tiny DB4}\\\hline
a_1,a_2&a_3,a_4&a_5,a_6&a_7,a_8\\
b_1,b_2&b_3,b_4&b_5,b_6&b_7,b_8\\\hline
\end{array}\cdots\\
&&\cdots\stackrel{(3)}{\longrightarrow}\begin{array}{|c|c|c|c|c|}\hline
\mbox{\tiny DB1}&\mbox{\tiny DB2}&\mbox{\tiny DB3}&\mbox{\tiny DB4}\\\hline
a_1,a_2&a_3,a_4&a_5,a_6&a_7,a_8\\
b_1,b_2&b_3,b_4&b_5,b_6&b_7,b_8\\
a_9+b_9&a_{11}+b_{11}&&\\
a_{10}+b_{10}&a_{12}+b_{12}&&\\
\hline
\end{array}
\stackrel{(1)}{\longrightarrow}\begin{array}{|c|c|c|c|c|}\hline
\mbox{\tiny DB1}&\mbox{\tiny DB2}&\mbox{\tiny DB3}&\mbox{\tiny DB4}\\\hline
a_1,a_2&a_3,a_4&a_5,a_6&a_7,a_8\\
b_1,b_2&b_3,b_4&b_5,b_6&b_7,b_8\\
a_9+b_9&a_{11}+b_{11}&a_{13}+b_{13}&a_{15}+b_{15}\\
a_{10}+b_{10}&a_{12}+b_{12}&a_{14}+b_{14}&a_{16}+b_{16}\\
\hline
\end{array}
\end{eqnarray*}
As before, we start by requesting $T^{K-1}=2$ symbols of desired message from each of the first $T=2$ databases. After enforcing symmetries across databases and then across messages, we consider the first $T$ databases (DB1 and DB2) and measure the total amount of external side information available outside DB1 and DB2, which turns out to be $4$ symbols ($b_5, b_6, b_7, b_8$). This gives us our budget of $2$ symbols of side information per database to be exploited to retrieve new desired information symbols. Assigning new labels to the side information symbols being exploited by each database, we add $a_9+b_9, a_{10}+b_{10}$ to the query from DB1 and $a_{11}+b_{11}, a_{12}+b_{12}$ to the query from DB2. Finally, enforcing symmetry across databases, corresponding queries are added to DB3 and DB4. At this point all symmetries are satisfied and the amounts of side-information available and exploited are balanced. Thus the structure is complete.

\subsubsection{Specialization}
Let each message consist of $N^K=16$ symbols from a sufficiently large finite field $\mathbb{F}_q$. The messages  $W_1$, $W_2\in\mathbb{F}_q^{16\times 1}$ are therefore represented as $16\times 1$ vectors over $\mathbb{F}_q$. Let $S_1, S_2\in\mathbb{F}_q^{16\times 16}$ represent random matrices chosen privately by the user, independently and uniformly from all $16\times 16$  full-rank matrices over $\mathbb{F}_q$. Without loss of generality, let us assume that $W_1$ is the desired message. Define $16\times 1$ vectors $a_{[1:16]}\in \mathbb{F}_q^{16\times 1}$ and $b_{[1:16]}\in \mathbb{F}_q^{16 \times 1}$, as follows
\begin{eqnarray}
a_{[1:16]}&=&S_1W_1\\
b_{[1:16]}&=&\mbox{MDS}_{16\times 8}S_2[(1:8),:]W_2
\end{eqnarray}
where $S_2[(1:8),:]$ is a $8\times 16$ matrix comprised of the first $8$ rows of $S_2$. MDS$_{16\times 8}$ is the generator matrix of a $(16, 8)$ MDS code. Plugging into the query structure obtained above,\begin{eqnarray*}
&&\begin{array}{|c|c|c|c|c|}\hline
\mbox{\tiny DB1}&\mbox{\tiny DB2}&\mbox{\tiny DB3}&\mbox{\tiny DB4}\\\hline
a_1,a_2&a_3,a_4&a_5,a_6&a_7,a_8\\
b_1,b_2&b_3,b_4&b_5,b_6&b_7,b_8\\
a_9+b_9&a_{11}+b_{11}&a_{13}+b_{13}&a_{15}+b_{15}\\
a_{10}+b_{10}&a_{12}+b_{12}&a_{14}+b_{14}&a_{16}+b_{16}\\
\hline
\end{array}
\end{eqnarray*}
the proof of correctness and privacy follows from the same reasoning as in the previous example. Note that since $16$ desired symbols are recovered from a total of $24$ downloaded symbols, the rate achieved by this scheme is $16/24=2/3$, which matches the capacity for this setting.

\subsection{$K = 2$ Messages, $N = 4$ Databases, $T = 3$ Colluding Databases}
The capacity for this setting, is $C=\left(1+\frac{3}{4}\right)^{-1}=\frac{4}{7}$.

\subsubsection{Query Structure}
The query structure is constructed as follows.
\begin{eqnarray*}
&&\begin{array}{|c|c|c|c|c|c|}\hline
\mbox{\tiny DB1}&\mbox{\tiny DB2}&\mbox{\tiny DB3}&\mbox{\tiny DB4}\\\hline
a_1,a_2, a_3&a_4,a_5,a_6&a_7,a_8,a_9&~~\\\hline
\end{array}\stackrel{(1)}{\longrightarrow}\begin{array}{|c|c|c|c|c|}\hline
\mbox{\tiny DB1}&\mbox{\tiny DB2}&\mbox{\tiny DB3}&\mbox{\tiny DB4}\\\hline
a_1,a_2,a_3&a_4,a_5,a_6&a_7,a_8,a_9&a_{10},a_{11},a_{12}\\\hline
\end{array}\cdots\\
&&\cdots \stackrel{(2)}{\longrightarrow}\begin{array}{|c|c|c|c|c|}\hline
\mbox{\tiny DB1}&\mbox{\tiny DB2}&\mbox{\tiny DB3}&\mbox{\tiny DB4}\\\hline
a_1,a_2,a_3&a_4,a_5,a_6&a_7,a_8,a_9&a_{10},a_{11},a_{12}\\
b_1,b_2,b_3&b_4,b_5,b_6&b_7,b_8,b_9&b_{10},b_{11},b_{12}\\\hline
\end{array}\cdots\\
&&\cdots\stackrel{(3)}{\longrightarrow}\begin{array}{|c|c|c|c|c|}\hline
\mbox{\tiny DB1}&\mbox{\tiny DB2}&\mbox{\tiny DB3}&\mbox{\tiny DB4}\\\hline
a_1,a_2,a_3&a_4,a_5,a_6&a_7,a_8,a_9&a_{10},a_{11},a_{12}\\
b_1,b_2,b_3&b_4,b_5,b_6&b_7,b_8,b_9&b_{10},b_{11},b_{12}\\
a_{13}+b_{13}&a_{14}+b_{14}&a_{15}+b_{15}&\\
\hline
\end{array}\cdots\\
&&\cdots
\stackrel{(1)}{\longrightarrow}\begin{array}{|c|c|c|c|c|}\hline
\mbox{\tiny DB1}&\mbox{\tiny DB2}&\mbox{\tiny DB3}&\mbox{\tiny DB4}\\\hline
a_1,a_2,a_3&a_4,a_5,a_6&a_7,a_8,a_9&a_{10},a_{11},a_{12}\\
b_1,b_2,b_3&b_4,b_5,b_6&b_7,b_8,b_9&b_{10},b_{11},b_{12}\\
a_{13}+b_{13}&a_{14}+b_{14}&a_{15}+b_{15}&a_{16}+b_{16}\\
\hline
\end{array}
\end{eqnarray*}
Starting with $T^{K-1}=3$ symbols each requested from the first $T=3$ databases, after the enforcing of symmetries across databases and then across messages, we jointly consider the first $T=3$ databases, DB1, DB2 and DB3. The amount of external side information available is $3$ symbols ($b_{10}, b_{11}, b_{12}$), which allows a budget of one symbol of side information per database to be exploited to recover new symbols of desired information. Assigning new labels to the side information symbols being exploited by each database, we include $a_{13}+b_{13}, a_{14}+b_{14}, a_{15}+b_{15}$ in the queries from DB1, DB2 and DB3, respectively. Finally, enforcing symmetry across databases, correspondingly $a_{16}+b_{16}$ is added to DB4. At this point all symmetries are satisfied and the amounts of side-information available and exploited are balanced. Thus the structure is complete.
\subsubsection{Specialization}
Let each message consist of $N^K=16$ symbols from a sufficiently large finite field $\mathbb{F}_q$. The messages  $W_1$, $W_2\in\mathbb{F}_q^{16\times 1}$ are therefore represented as $16\times 1$ vectors over $\mathbb{F}_q$. Let $S_1, S_2\in\mathbb{F}_q^{16\times 16}$ represent random matrices chosen privately by the user, independently and uniformly from all $16\times 16$  full-rank matrices over $\mathbb{F}_q$. Without loss of generality, let us assume that $W_1$ is the desired message. Define $16\times 1$ vectors $a_{[1:16]}\in \mathbb{F}_q^{16\times 1}$ and $b_{[1:16]}\in \mathbb{F}_q^{16 \times 1}$, as follows
\begin{eqnarray}
a_{[1:16]}&=&S_1W_1\\
b_{[1:16]}&=&\mbox{MDS}_{16\times 12}S_2[(1:12),:]W_2
\end{eqnarray}
where $S_2[(1:12),:]$ is a $12\times 16$ matrix comprised of the first $12$ rows of $S_2$. MDS$_{16\times 12}$ is the generator matrix of a $(16, 12)$ MDS code. After plugging these values into query structure derived earlier, the proof of correctness and privacy follows from the same reasoning as in the previous examples. Note that since $16$ desired symbols are recovered from a total of $28$ downloaded symbols, the rate achieved by this scheme is $16/28=4/7$, which matches the capacity for this setting.

\subsection{$K = 3$ Messages, $N = 3$ Databases, $T = 2$ Colluding Databases}
The capacity for this setting, is $C=\left(1+\frac{2}{3}+(\frac{2}{3})^2\right)^{-1}=\frac{9}{19}$.
\subsubsection{Query Structure}
The query structure is constructed as follows.
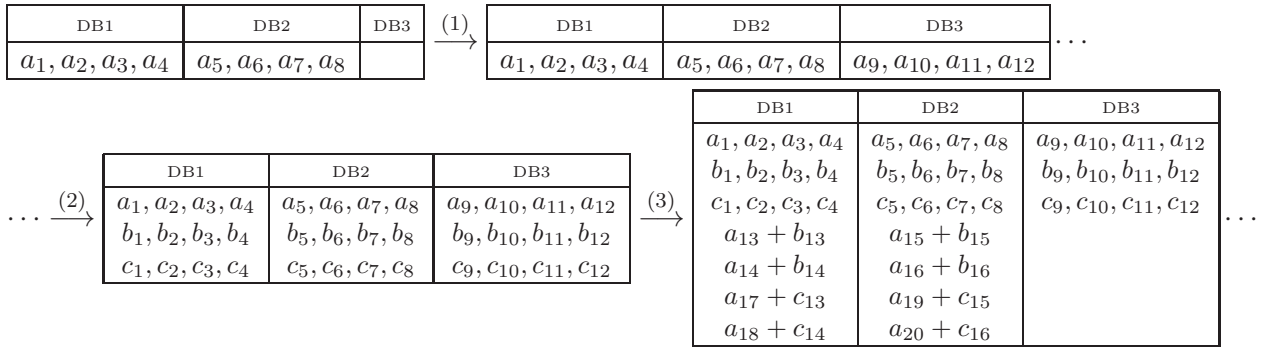
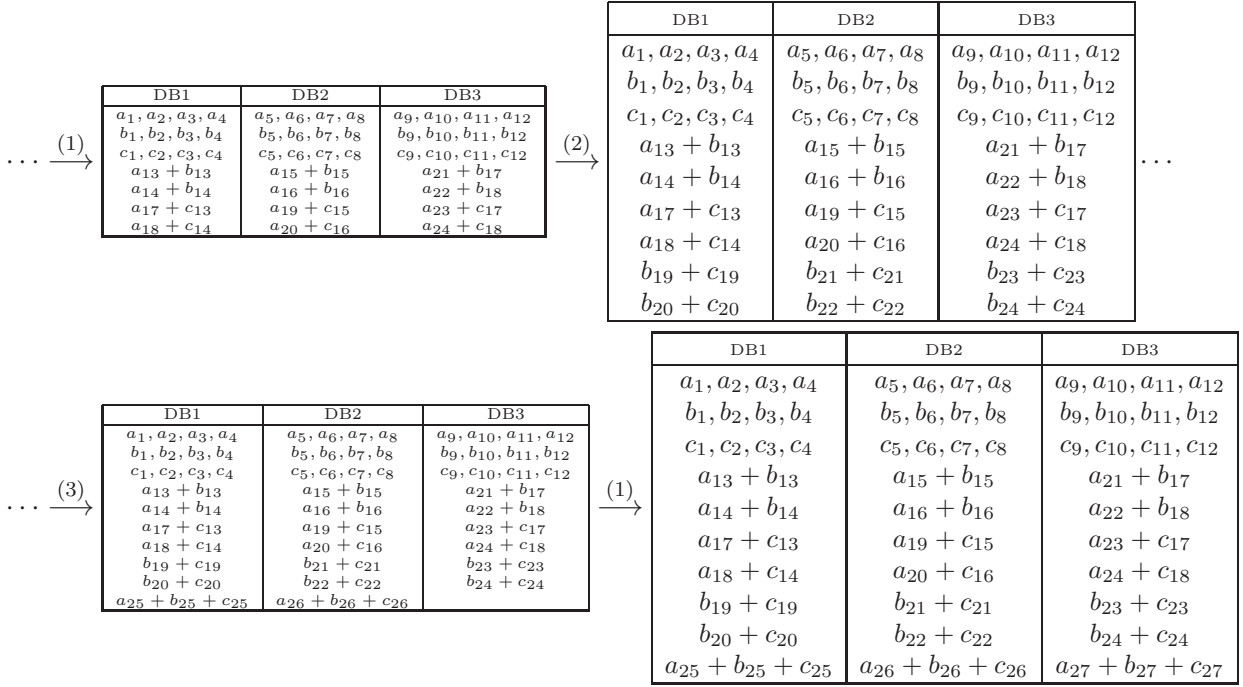
\begin{eqnarray*}
&&\begin{array}{|c|c|c|c|c|}\hline
\mbox{\tiny DB1}&\mbox{\tiny DB2}&\mbox{\tiny DB3}\\\hline
a_1,a_2,a_3,a_4&a_5,a_6,a_7,a_8&~~\\\hline
\end{array}\stackrel{(1)}{\longrightarrow}\begin{array}{|c|c|c|c|c|}\hline
\mbox{\tiny DB1}&\mbox{\tiny DB2}&\mbox{\tiny DB3}\\\hline
a_1,a_2,a_3,a_4&a_5,a_6,a_7,a_8&a_9,a_{10},a_{11},a_{12}\\\hline
\end{array}\cdots\\
&&\cdots
\stackrel{(2)}{\longrightarrow}{\small\begin{array}{|c|c|c|c|c|}\hline
\mbox{\tiny DB1}&\mbox{\tiny DB2}&\mbox{\tiny DB3}\\\hline
a_1,a_2,a_3,a_4&a_5,a_6,a_7,a_8&a_9,a_{10},a_{11},a_{12}\\
b_1,b_2,b_3,b_4&b_5,b_6,b_7,b_8&b_9,b_{10},b_{11},b_{12}\\
c_1,c_2,c_3,c_4&c_5,c_6,c_7,c_8&c_9,c_{10},c_{11},c_{12}\\
\hline
\end{array}}
\stackrel{(3)}{\longrightarrow}{\small\begin{array}{|c|c|c|c|c|}\hline
\mbox{\tiny DB1}&\mbox{\tiny DB2}&\mbox{\tiny DB3}\\\hline
a_1,a_2,a_3,a_4&a_5,a_6,a_7,a_8&a_9,a_{10},a_{11},a_{12}\\
b_1,b_2,b_3,b_4&b_5,b_6,b_7,b_8&b_9,b_{10},b_{11},b_{12}\\
c_1,c_2,c_3,c_4&c_5,c_6,c_7,c_8&c_9,c_{10},c_{11},c_{12}\\
a_{13}+b_{13}&a_{15}+b_{15}&\\
a_{14}+b_{14}&a_{16}+b_{16}&\\
a_{17}+c_{13}&a_{19}+c_{15}&\\
a_{18}+c_{14}&a_{20}+c_{16}&\\
\hline
\end{array}}
\cdots\\
&&\cdots
\stackrel{(1)}{\longrightarrow}
{\tiny\begin{array}{|c|c|c|c|c|}\hline
\mbox{\tiny DB1}&\mbox{\tiny DB2}&\mbox{\tiny DB3}\\\hline
a_1,a_2,a_3,a_4&a_5,a_6,a_7,a_8&a_9,a_{10},a_{11},a_{12}\\
b_1,b_2,b_3,b_4&b_5,b_6,b_7,b_8&b_9,b_{10},b_{11},b_{12}\\
c_1,c_2,c_3,c_4&c_5,c_6,c_7,c_8&c_9,c_{10},c_{11},c_{12}\\
a_{13}+b_{13}&a_{15}+b_{15}&a_{21}+b_{17}\\
a_{14}+b_{14}&a_{16}+b_{16}&a_{22}+b_{18}\\
a_{17}+c_{13}&a_{19}+c_{15}&a_{23}+c_{17}\\
a_{18}+c_{14}&a_{20}+c_{16}&a_{24}+c_{18}\\
\hline
\end{array}}
\stackrel{(2)}{\longrightarrow}
{\small\begin{array}{|c|c|c|c|c|}\hline
\mbox{\tiny DB1}&\mbox{\tiny DB2}&\mbox{\tiny DB3}\\\hline
a_1,a_2,a_3,a_4&a_5,a_6,a_7,a_8&a_9,a_{10},a_{11},a_{12}\\
b_1,b_2,b_3,b_4&b_5,b_6,b_7,b_8&b_9,b_{10},b_{11},b_{12}\\
c_1,c_2,c_3,c_4&c_5,c_6,c_7,c_8&c_9,c_{10},c_{11},c_{12}\\
a_{13}+b_{13}&a_{15}+b_{15}&a_{21}+b_{17}\\
a_{14}+b_{14}&a_{16}+b_{16}&a_{22}+b_{18}\\
a_{17}+c_{13}&a_{19}+c_{15}&a_{23}+c_{17}\\
a_{18}+c_{14}&a_{20}+c_{16}&a_{24}+c_{18}\\
b_{19}+c_{19}&b_{21}+c_{21}&b_{23}+c_{23}\\
b_{20}+c_{20}&b_{22}+c_{22}&b_{24}+c_{24}\\
\hline
\end{array}}
\cdots\\
&&\cdots
\stackrel{(3)}{\longrightarrow}
{\tiny \begin{array}{|c|c|c|c|c|}\hline
\mbox{\tiny DB1}&\mbox{\tiny DB2}&\mbox{\tiny DB3}\\\hline
a_1,a_2,a_3,a_4&a_5,a_6,a_7,a_8&a_9,a_{10},a_{11},a_{12}\\
b_1,b_2,b_3,b_4&b_5,b_6,b_7,b_8&b_9,b_{10},b_{11},b_{12}\\
c_1,c_2,c_3,c_4&c_5,c_6,c_7,c_8&c_9,c_{10},c_{11},c_{12}\\
a_{13}+b_{13}&a_{15}+b_{15}&a_{21}+b_{17}\\
a_{14}+b_{14}&a_{16}+b_{16}&a_{22}+b_{18}\\
a_{17}+c_{13}&a_{19}+c_{15}&a_{23}+c_{17}\\
a_{18}+c_{14}&a_{20}+c_{16}&a_{24}+c_{18}\\
b_{19}+c_{19}&b_{21}+c_{21}&b_{23}+c_{23}\\
b_{20}+c_{20}&b_{22}+c_{22}&b_{24}+c_{24}\\
a_{25}+b_{25}+c_{25}&a_{26}+b_{26}+c_{26}&\\
\hline
\end{array}}
%\\&&\cdots
 \stackrel{(1)}{\longrightarrow}
 {\small\begin{array}{|c|c|c|c|c|}\hline
\mbox{\tiny DB1}&\mbox{\tiny DB2}&\mbox{\tiny DB3}\\\hline
a_1,a_2,a_3,a_4&a_5,a_6,a_7,a_8&a_9,a_{10},a_{11},a_{12}\\
b_1,b_2,b_3,b_4&b_5,b_6,b_7,b_8&b_9,b_{10},b_{11},b_{12}\\
c_1,c_2,c_3,c_4&c_5,c_6,c_7,c_8&c_9,c_{10},c_{11},c_{12}\\
a_{13}+b_{13}&a_{15}+b_{15}&a_{21}+b_{17}\\
a_{14}+b_{14}&a_{16}+b_{16}&a_{22}+b_{18}\\
a_{17}+c_{13}&a_{19}+c_{15}&a_{23}+c_{17}\\
a_{18}+c_{14}&a_{20}+c_{16}&a_{24}+c_{18}\\
b_{19}+c_{19}&b_{21}+c_{21}&b_{23}+c_{23}\\
b_{20}+c_{20}&b_{22}+c_{22}&b_{24}+c_{24}\\
a_{25}+b_{25}+c_{25}&a_{26}+b_{26}+c_{26}&a_{27}+b_{27}+c_{27}\\
\hline
\end{array}}
\end{eqnarray*}
Starting with $T^{K-1}=4$ symbols each requested from the first $T=2$ databases, we proceed through iterative steps (1) and (2) to enforce symmetries across databases and messages. In step (3) we consider the first $T=2$ databases together (DB1 and DB2) and account for the external side information, which in this case contains $4$ symbols from $[b_k]$  and $4$ symbols from $[c_k]$. Distributed evenly among DB1 and DB2, this allows a budget of $2$ symbols of side information from $[b_k]$ and 2 symbols from $[c_k]$ per database to be exploited to recover new  symbols of desired information. Proceeding again through steps (1) and (2) to enforce symmetries across databases and messages, we end up with new downloads that contain only undesired information symbols, which can now be used to download new desired information symbols. Once again, we consider DB1 and DB2 together, and account for the new external side information, $b_{23}+c_{23}, b_{24}+c_{24}$. Thus the external side information is comprised of two symbols, each of which is a sum of the form $b_k+c_k$. Dividing the side information evenly among databases DB1 and DB2, each is assigned one side-information symbol of the form $b_k+c_k$ with new labels. Thus, $a_{25}+b_{25}+c_{25}$ is added to the query from DB1, and $a_{26}+b_{26}+c_{26}$ is added to the query from DB2. Finally, applying symmetry across databases, we include $a_{27}+b_{27}+c_{27}$ to the query from DB3. At this point, all symmetries are satisfied, all external and exploited side-information amounts are balanced, and therefore, the query structure is complete.

\subsubsection{Specialization}
Let each message consist of $N^K=27$ symbols from a sufficiently large finite field $\mathbb{F}_q$. The messages  $W_1,W_2,W_3\in\mathbb{F}_q^{27\times 1}$ are then represented as $27\times 1$ vectors over $\mathbb{F}_q$. Let $S_1, S_2,S_3\in\mathbb{F}_q^{27\times 27}$ represent random matrices chosen privately by the user, independently and uniformly from all $27\times 27$  full-rank matrices over $\mathbb{F}_q$. Without loss of generality, let us assume that $W_1$ is the desired message. Define $27\times 1$ vectors $a_{[1:27]},b_{[1:27]},c_{[1:27]}\in \mathbb{F}_q^{27 \times 1}$, as follows
\begin{eqnarray}
a_{[1:27]}&=&S_1W_1\\
b_{[1:18]}&=&\mbox{MDS}_{18\times 12}S_2[(1:12),:]W_2\label{eq:g1}\\
c_{[1:18]}&=&\mbox{MDS}_{18\times 12}S_3[(1:12),:]W_3\label{eq:g2}\\
b_{[19:27]}&=&\mbox{MDS}_{9\times 6}S_2[(13:18),:]W_2\label{eq:g3}\\
c_{[19:27]}&=&\mbox{MDS}_{9\times 6}S_3[(13:18),:]W_3\label{eq:g4}
\end{eqnarray}
where $S_2[(1:18),:]$ is a $18\times 27$ matrix comprised of the first $18$ rows of $S_2$. MDS$_{18\times 12}$ is the generator matrix of a $(18, 12)$ MDS code, and MDS$_{9\times 6}$ is the generator matrix of a $(9, 6)$ MDS code. In particular, note that the \emph{same} generator matrix is used in (\ref{eq:g1}) and (\ref{eq:g2}). Similarly, the same generator matrix is used in (\ref{eq:g3}) and (\ref{eq:g4}). This is important because it allows us to write
\begin{eqnarray}
b_{[19:27]}+c_{[19:27]}&=&\mbox{MDS}_{9\times 6}\left(S_2[(13:18),:]W_2+S_3[(13:18),:]W_3\right) \label{align}
\end{eqnarray}
so that all $9$ elements of the vector $b_{[19:27]}+c_{[19:27]}$ can be recovered from any $6$ of its elements, e.g., from $b_{[19:24]}+c_{[19:24]}$ one can also recover $b_{25}+c_{25}, b_{26}+c_{26}, b_{27}+c_{27}$. This observation is the key to understanding the role of interference alignment in this construction. The effective number of \emph{resolvable} undesired symbols is minimized due to interference alignment. For example, $b_{19}$ and $c_{19}$ are always aligned together into one symbol $b_{19}+c_{19}$ in all the downloaded equations. The two are unresolvable from each other and act as effectively one undesired symbol in the downloaded equations, thus reducing the effective number of undesired symbols, so that the same number of downloaded equations can be used to retrieve a greater number of desired symbols. Note also that desired symbols are always resolvable.

These values are plugged into the query structure derived previously.
\begin{eqnarray*}
 {\small\begin{array}{|c|c|c|c|c|}\hline
\mbox{\tiny DB1}&\mbox{\tiny DB2}&\mbox{\tiny DB3}\\\hline
a_1,a_2,a_3,a_4&a_5,a_6,a_7,a_8&a_9,a_{10},a_{11},a_{12}\\
b_1,b_2,b_3,b_4&b_5,b_6,b_7,b_8&b_9,b_{10},b_{11},b_{12}\\
c_1,c_2,c_3,c_4&c_5,c_6,c_7,c_8&c_9,c_{10},c_{11},c_{12}\\
a_{13}+b_{13}&a_{15}+b_{15}&a_{21}+b_{17}\\
a_{14}+b_{14}&a_{16}+b_{16}&a_{22}+b_{18}\\
a_{17}+c_{13}&a_{19}+c_{15}&a_{23}+c_{17}\\
a_{18}+c_{14}&a_{20}+c_{16}&a_{24}+c_{18}\\
b_{19}+c_{19}&b_{21}+c_{21}&b_{23}+c_{23}\\
b_{20}+c_{20}&b_{22}+c_{22}&b_{24}+c_{24}\\
a_{25}+b_{25}+c_{25}&a_{26}+b_{26}+c_{26}&a_{27}+b_{27}+c_{27}\\
\hline
\end{array}}\end{eqnarray*}
Correctness is straightforward. Let us see why $T$-privacy holds.  The queries for any $T=2$ colluding databases are comprised of $18$ variables from $a_{[1:27]}$, $12$ variables from $b_{[1:18]}$, $6$ variables from $b_{[19:27]}$, $12$ variables from $c_{[1:18]}$ and $6$ variables from $c_{[19:27]}$. Let the indices of these variables be denoted by the  vectors $\mathcal{I}_{a}\in\mathbb{N}^{18\times 1}, \mathcal{I}_{b,12}\in\mathbb{N}^{12\times 1},\mathcal{I}_{b,6}\in\mathbb{N}^{6\times 1},\mathcal{I}_{c,12}\in\mathbb{N}^{12\times 1}$ and $\mathcal{I}_{c,6}\in\mathbb{N}^{6\times 1}$, respectively, so that,
\begin{eqnarray}
a_{\mathcal{I}_a}&=&S_1[\mathcal{I}_a,:]W_1\\
b_{\mathcal{I}_{b,12}}&=&\mbox{MDS}_{18\times 12}[\mathcal{I}_{b,12},:]S_2[(1:12),:]W_2\\
b_{\mathcal{I}_{b,6}}&=&\mbox{MDS}_{9\times 6}[\mathcal{I}_{b,6},:]S_2[(13:18),:]W_2\\
c_{\mathcal{I}_{c,12}}&=&\mbox{MDS}_{18\times 12}[\mathcal{I}_{c,12},:]S_3[(1:12),:]W_3\\
c_{\mathcal{I}_{c,6}}&=&\mbox{MDS}_{9\times 6}[\mathcal{I}_{c,6},:]S_3[(13:18),:]W_3
\end{eqnarray}
From Lemma \ref{lemma:inv}, we have
\begin{eqnarray}
(a_{\mathcal{I}_a}, (b_{\mathcal{I}_{b,12}};~~ b_{\mathcal{I}_{b,6}}), (c_{\mathcal{I}_{c,12}}; ~~c_{\mathcal{I}_{c,6}}))&\sim& (S_1[(1:18),:]W_1,S_2[(1:18),:]W_2,S_3[(1:18),:]W_3)
\end{eqnarray}
Thus privacy is guaranteed. Finally, note that since $27$ desired symbols are recovered from a total of $57$ downloaded symbols, the rate achieved by this scheme is $27/57=9/19$, which matches the capacity for this setting.

\subsection{Arbitrary Number of Messages $K$, Arbitrary Number of Databases $N$, Arbitrary Number of Colluding Databases $T$}
\subsubsection{Query Structure}
For arbitrary $K, N, T$, we follow the same iterative procedure, briefly summarized below.
\begin{itemize}
\item Step 1: Initialization. Download $T^{K-1}$ desired symbols each from the first $T$ databases.
\item Step 2: Invoke symmetry across databases to determine corresponding downloads from DB $T+1$ to DB $N$.
\item Step 3: Invoke symmetry of messages to determine additional downloaded equations (comprised only of undesired symbols) from each database.
\item Step 4: Consider the first $T$ databases together. Divide the new external side information generated in the previous step evenly among the first $T$ databases to determine the side-information budget per database. For each side information symbol allocated to a database create an additional query of the same form as the assigned side information (with new labels) combined with a new desired symbol. 
\item Step 5: Go back to Step 2 and run Step 2 to Step 4 a total of $(K-1)$ times.
\end{itemize}

\subsubsection{Specialization}
{\color{black} Let each message consist of $N^K$ symbols from a sufficiently large finite field $\mathbb{F}_q$. The messages  $W_1,\cdots,W_K\in\mathbb{F}_q^{N^K\times 1}$ are  represented as $N^K\times 1$ vectors over $\mathbb{F}_q$. Let $S_1, \cdots, S_K\in\mathbb{F}_q^{N^K\times N^K}$ represent random matrices chosen privately by the user, independently and uniformly from all $N^K\times N^K$  full-rank matrices over $\mathbb{F}_q$. Suppose $W_l$, $l\in[1:K]$, is the desired message. 

Consider any undesired message index $k\in [1:K]/\{l\}$, and all distinct $\Delta=2^{K-2}$  subsets of $[1:K]$ that contain $k$ and do not contain $l$. Assign distinct labels to each subset, e.g., $\mathcal{K}_1, \mathcal{K}_2, \cdots \mathcal{K}_{\Delta}$. 
For each $k\in [1:K]/\{l\}$, define the vector

{\setstretch{1.5}
\begin{eqnarray*}
\left[\begin{array}{l}x^{[k]}_{\mathcal{K}_1}\\
x^{[k]}_{\mathcal{K}_1\cup\{l\}}\\ \hdashline[2pt/2pt]
x^{[k]}_{\mathcal{K}_2}\\
x^{[k]}_{\mathcal{K}_2\cup\{l\}}\\ \hdashline[2pt/2pt]
\vdots\\ \hdashline[2pt/2pt]
x^{[k]}_{\mathcal{K}_\Delta}\\
x^{[k]}_{\mathcal{K}_\Delta\cup\{l\}}\\
\end{array}
\right]&=&\left[\begin{array}{llll}
\mbox{MDS}_{\frac{N}{T}\alpha_1\times \alpha_1}&0&0&0\\ \hdashline[2pt/2pt]
0&\mbox{MDS}_{\frac{N}{T}\alpha_2\times \alpha_2}&0&0\\ \hdashline[2pt/2pt]
0&\cdots&\ddots&0\\ \hdashline[2pt/2pt]
0&0&0&\mbox{MDS}_{\frac{N}{T}\alpha_\Delta\times \alpha_\Delta}\\
\end{array}\right]S_k[(1:TN^{K-1}),:]W_k
\end{eqnarray*}
}
\noindent where $\alpha_i, i \in [1:\Delta]$ is defined as $N(N-T)^{|\mathcal{K}_i| - 1} T^{K - |\mathcal{K}_i|}$,  each $x^{[k]}_{\mathcal{K}_i}$ is a $\alpha_i \times 1$ vector, and each $x^{[k]}_{\mathcal{K}_i\cup\{l\}}$ is a $(\frac{N-T}{T})\alpha_i \times 1$ vector over $\mathbb{F}_q$.

Now consider the desired message index $l$, and all distinct $\delta=2^{K-1}$ subsets of $[1:K]$ that contain $l$. Assign distinct labels to each subset, e.g., $\mathcal{L}_1, \mathcal{L}_2, \cdots \mathcal{L}_{\delta}$. Define the vector
{\setstretch{1.5}
\begin{eqnarray*}
\left[\begin{array}{l}x^{[l]}_{\mathcal{L}_1}\\
x^{[l]}_{\mathcal{L}_2}\\
\vdots\\
x^{[l]}_{\mathcal{L}_\delta}\\
\end{array}
\right]&=&S_lW_l
\end{eqnarray*}
where the length of $x_{\mathcal{L}_i}^{[l]}, i \in [1:\delta]$ is $N(N-T)^{|\mathcal{L}_i| - 1} T^{K - |\mathcal{L}_i|}$.

For each non-empty subset $\mathcal{K}\subset[1:K]$ generate the query vector
\begin{eqnarray}
\sum_{k\in\mathcal{K}}x^{[k]}_\mathcal{K}
\end{eqnarray}
Distribute the elements of the query vector evenly among the $N$ databases. This completes the specialized construction of the queries.
}}

The construction has $K$ layers. Over the $j$-th layer, from each database, we download  $(N-T)^{j-1}T^{K-j} \binom{K}{j}$ equations that are comprised of sums of $j$ symbols, out of which $(N-T)^{j-1}T^{K-j} \binom{K-1}{j-1}$ involve desired data symbols. Our construction ensures that the interference $x^{[k]}_{\mathcal{K}_i \cup \{l\}}, k \neq l, i \in [1:\Delta]$ in the $(|\mathcal{K}_i| + 1)$-th layer can be recovered from the corresponding symbols $x^{[k]}_{\mathcal{K}_i}$ in the $|\mathcal{K}_i|$-th layer.
Therefore correctness is guaranteed.

Let us see why privacy holds. The queries for any $T$ colluding databases are comprised of $TN^{K-1}$ variables from each $x^{[k]}, k \in [1:K]$. In particular, $\forall k \neq l$, the variables from $x^{[k]}$ consist of $\alpha_i$ variables out of $\frac{N}{T}\alpha_i$  variables $x_{\mathcal{K}_i}^{[k]}, x_{\mathcal{K}_i\cup\{l\}}^{[k]}$, for each set $\mathcal{K}_i, i\in[1:\Delta]$. Note that these $\alpha_i$ variables are generated by the generator matrix of a $(\frac{N}{T}\alpha_{i}, \alpha_{i})$ MDS code, so that they have full rank.
Let the indices of the appeared variables be denoted by the  vectors $\mathcal{I}_{x^{[k]}}\in\mathbb{N}^{TN^{K-1}\times 1}, \forall k \in [1:K]$.
From Lemma \ref{lemma:inv}, we have
\begin{eqnarray}
%x^{[l]}_{\mathcal{I}_{x^{[l]}}} &\sim& S_l[\mathcal{I}_{x^{[l]}},:]W_l\\
x^{[k]}_{\mathcal{I}_{x^{[k]}}} &\sim& S_k[(1:TN^{K-1}),:]W_k%, k\neq l
\end{eqnarray}
which  in turn implies that %$S_l[\mathcal{I}_{x^{[l]}},:]$, 
$S_k[(1:TN^{K-1}),:]%, k\neq l
$ are independent and identically distributed. Thus privacy is guaranteed.

Finally, we compute the ratio of the number of desired symbols to the number of total downloaded symbols,
\begin{eqnarray}
{R} 
&=& \frac{N}{N} \frac{T^{K-1}\binom{K-1}{0} + (N-T) T^{K-2}\binom{K-1}{1} + (N-T)^2 T^{K-3} \binom{K-1}{2} + \cdots + (N-T)^{K-1}\binom{K-1}{K-1}}{T^{K-1}\binom{K}{1} + (N-T) T^{K-2}\binom{K}{2} + (N-T)^2 T^{K-3} \binom{K}{3} + \cdots +(N-T)^{K-1} \binom{K}{K}} \notag\\
&&\\
&=& \frac{N}{N} \frac{N^{K-1}} {\frac{1}{N-T} \left[ (N-T)T^{K-1}\binom{K}{1} + (N-T)^2T^{K-2}\binom{K}{2} + \cdots + (N-T)^{K}  \binom{K}{K} \right] }
\\
&=&\frac{\frac{1}{N}N^{K}}{\frac{1}{N-T} \left( N^{K} - T^{K} \right) } = \frac{1 - \frac{T}{N}}{1 - \frac{T^{K}}{N^{K}}} \\
&=& \left(1 + \frac{T}{N} + \frac{T^2}{N^2} + \cdots + \frac{T^{K-1}}{N^{K-1}}\right)^{-1}
\end{eqnarray}
Thus, the PIR rate achieved by the scheme always matches the capacity.

\section{Proof of Theorem \ref{thm:download}: Converse}\label{sec:con}
For compact notation, let us define 
\begin{eqnarray}
\mathcal{Q}&\define&\{Q_{n}^{[k]}: k\in[1:K], n\in[1:N]\}\\
A_{\mathcal{I}}^{[k]} &\define& \{A_{n}^{[k]}: n\in\mathcal{I} \}\\
\mathcal{H}_{T}&\define& \frac{1}{\binom{N}{T}} \sum_{\mathcal{T}:|\mathcal{T}| = T} \frac{H(A_{\mathcal{T}}|\mathcal{Q})}{T}, \mathcal{T} \subset [1:N]
\end{eqnarray}

We first state Han's inequality (Theorem 17.6.1 in \cite{Cover_Thomas}), which will be used later and is described here for the sake of completeness.

\begin{theorem}\label{thm:han} (Han's inequality)
\begin{eqnarray}
\mathcal{H}_{T} \geq \frac{H(A_{1}^{[k]}, A_{2}^{[k]}, \cdots, A_{N}^{[k]}|\mathcal{Q})}{N}
\end{eqnarray}
\end{theorem}

We next proceed to the converse proof.
The proof of outer bound for Theorem \ref{thm:download} is based on an induction argument. To set up the induction, we will prove the outer bound for $K=1$ and for $K=2$, each for arbitrary $N, T$, and then proceed to the case of arbitrary $K$.

\subsection{$K=1$ Message, $N$ Databases}
%This case is straightforward, because if there is only one message, then the query is trivially independent of the message index. The user only needs to download his desired message bits from each database. As expected, the rate must be at most $1$.
\begin{eqnarray}
L =  H(W_1) 
&=& H(W_1|\mathcal{Q}) \\
&\leq&I(A_1^{[1]}, A_2^{[1]}, \cdots, A_N^{[1]};W_1|\mathcal{Q}) \\
&=&H(A_1^{[1]}, A_2^{[1]}, \cdots, A_N^{[1]}|\mathcal{Q}) \\
&\leq& N\mathcal{H}_T \label{eq:ind1}\\
&\leq& \sum_{n = 1}^N H(A_n|\mathcal{Q}) \label{eq:last}\\
\Rightarrow R&=&\frac{L}{\sum_{n = 1}^N H(A_n)}\leq \frac{L}{\sum_{n = 1}^N H(A_n|\mathcal{Q})} \leq 1
\end{eqnarray}
where (\ref{eq:ind1}) follows from Han's inequality, and (\ref{eq:last}) is due to the property that dropping conditioning does not reduce entropy.

\subsection{$K = 2$ Messages, $N$ Databases}
Consider  $\mathcal{T}\subset[1:N]$ with cardinality $|\mathcal{T}|=T$. Denote the complement of $\mathcal{T}$ as $\overline{\mathcal{T}}$.

Since the queries and answer strings for any $T$ databases are independent of the message index,  we will denote $A^{[k]}_\mathcal{T}$ simply as $A_\mathcal{T}, \forall k\in[1:K]$.  From $A_{\mathcal{T}}, A_{\overline{\mathcal{T}}}^{[1]}, A_{\overline{\mathcal{T}}}^{[2]},
\mathcal{Q}$, we can decode $W_1, W_2$. 
\begin{eqnarray}
2L  =  H(W_1,W_2) 
&=& H(W_1, W_2 | \mathcal{Q}) \\ 
 &\leq&I(A_{\mathcal{T}}, A_{\overline{\mathcal{T}}}^{[1]}, A_{\overline{\mathcal{T}}}^{[2]}; W_1, W_2|\mathcal{Q})\\ 
&=& H(A_{\mathcal{T}}, A_{\overline{\mathcal{T}}}^{[1]}, A_{\overline{\mathcal{T}}}^{[2]} | \mathcal{Q}) \label{oo1} \\
&=& H(A_{\mathcal{T}}, A_{\overline{\mathcal{T}}}^{[1]}| \mathcal{Q})  + H(A_{\overline{\mathcal{T}}}^{[2]} | A_{\mathcal{T}}, A_{\overline{\mathcal{T}}}^{[1]}, \mathcal{Q})  \\
&=& H(A_{\mathcal{T}}, A_{\overline{\mathcal{T}}}^{[1]}| \mathcal{Q})  + H(A_{\overline{\mathcal{T}}}^{[2]} | A_{\mathcal{T}}, A_{\overline{\mathcal{T}}}^{[1]}, W_1, \mathcal{Q}) \label{o2} \\
&\leq& N\mathcal{H}_T  + H(A_{\overline{\mathcal{T}}}^{[2]} | A_{\mathcal{T}}, W_1, \mathcal{Q}) \label{oo2} \\
&=&   N\mathcal{H}_T   + H(A_{[1:N]}^{[2]} | W_{1},\mathcal{Q}) - H(A_{\mathcal{T}} | W_{1},\mathcal{Q}) \label{oo22} \\
&=& N\mathcal{H}_T   +  L -  H(A_{\mathcal{T}} | W_{1},\mathcal{Q})  
\label{oo23}
\end{eqnarray}
where (\ref{oo1}) is due to the fact that the answering strings are deterministic functions of the messages and queries.  (\ref{o2}) is due to the fact that $W_1$ is a  function of $(A_{\mathcal{T}}, A_{\overline{\mathcal{T}}}^{[1]}, \mathcal{Q})$. In (\ref{oo23}), the second term follows from the fact that from $A_{[1:N]}^{[2]}$, we can decode $W_2$.

Consider (\ref{oo23}) for all subsets of $[1:N]$ that have exactly $T$ elements and average over all such subsets. We have
\begin{eqnarray}
2L &\leq& N\mathcal{H}_T   +  L -  \frac{1}{\binom{N}{T}} \sum_{\mathcal{T}:|\mathcal{T}| = T}H(A_{\mathcal{T}} | W_{1},\mathcal{Q}) \\
&\leq& N\mathcal{H}_T   +  L -  TH(A_{[1:N]}^{[2]}|W_1, \mathcal{Q})/N \label{p2_han}\\
&=& N\mathcal{H}_T   +  L -  TL/N \label{p2_dec}\\
\Rightarrow L\left(1+\frac{T}{N}\right)
&\leq& N\mathcal{H}_T  \leq {\sum_{n=1}^N H(A_n | \mathcal{Q})}  \label{re2} \\
\Rightarrow R&=&\frac{L}{\sum_{n=1}^N H(A_n)}\leq \frac{L}{\sum_{n=1}^N H(A_n | \mathcal{Q})} \leq \left(1 + \frac{T}{N} \right)^{-1}
\end{eqnarray}
where (\ref{p2_han}) follows from the conditional version of Han's inequality, and (\ref{p2_dec}) is due to the fact that from $A_{[1:N]}^{[2]}$, we can decode $W_2$.
The outer bound proof for $K=2$ messages setting is complete.

\subsection{$K\geq 3$ Messages,  $N$ Databases}
Consider  $\mathcal{T}\subset[1:N]$ with cardinality $|\mathcal{T}|=T$.
From $A_{\mathcal{T}}, A_{\overline{\mathcal{T}}}^{[1]}, \cdots, A_{\overline{\mathcal{T}}}^{[K]},\mathcal{Q}$, we can decode  all $K$ messages $W_1, \cdots, W_K$. 
\begin{eqnarray}
&&KL = H(W_1, \cdots, W_K | \mathcal{Q})
\\
&\leq&I(A_{\mathcal{T}}, A_{\overline{\mathcal{T}}}^{[1]}, \cdots, A_{\overline{\mathcal{T}}}^{[K]};W_1,\cdots,W_K|\mathcal{Q})\\
&=& H(A_{\mathcal{T}}, A_{\overline{\mathcal{T}}}^{[1]}, \cdots, A_{\overline{\mathcal{T}}}^{[K]} | \mathcal{Q}) \label{ook1} \\
&=& H(A_{\mathcal{T}}, A_{\overline{\mathcal{T}}}^{[1]} | \mathcal{Q})  + H(A_{\overline{\mathcal{T}}}^{[2]}, \cdots, A_{\overline{\mathcal{T}}}^{[K]} | A_{\mathcal{T}}, A_{\overline{\mathcal{T}}}^{[1]}, \mathcal{Q})  \\
&\leq& N\mathcal{H}_T +  H(A_{\overline{\mathcal{T}}}^{[2]}, \cdots, A_{\overline{\mathcal{T}}}^{[K]} | A_{\mathcal{T}}, A_{\overline{\mathcal{T}}}^{[1]}, W_1, \mathcal{Q})   \label{ok2} \\
&\leq& N\mathcal{H}_T +H(A_{\overline{\mathcal{T}}}^{[2]}  |  A_{\mathcal{T}}, W_1,\mathcal{Q}) + H(A_{\overline{\mathcal{T}}}^{[3]}, \cdots, A_{\overline{\mathcal{T}}}^{[K]} | A_{\mathcal{T}}, A_{\overline{\mathcal{T}}}^{[2]}, W_1,\mathcal{Q}) 
\label{oksym} \\
&\leq& N\mathcal{H}_T + \sum_{n \in \overline{\mathcal{T}}} H(A_{n}^{[2]}  | A_{\mathcal{T}}, W_1,\mathcal{Q}) + H(A_{\overline{\mathcal{T}}}^{[3]}, \cdots, A_{\overline{\mathcal{T}}}^{[K]} | A_{\mathcal{T}}, W_1, W_2, \mathcal{Q})   ~~~~\label{ok3} \\
&=& N\mathcal{H}_T+ \sum_{n \in \overline{\mathcal{T}}} H(A_{n}^{[2]}  |A_{\mathcal{T}}, W_1,\mathcal{Q}) + H(A_{\mathcal{T}}, A_{\overline{\mathcal{T}}}^{[3]}, \cdots, A_{\overline{\mathcal{T}}}^{[K]} | W_1, W_2, \mathcal{Q})  - H(A_{\mathcal{T}}|W_1,W_2,\mathcal{Q}) \notag \\
&&\\
&=& N\mathcal{H}_T + \sum_{n \in \overline{\mathcal{T}}} H(A_{n}^{[2]}  |A_{\mathcal{T}}, W_1,\mathcal{Q}) + (K-2)L  - H(A_{\mathcal{T}}|W_1,W_2,\mathcal{Q}) \label{pk_1}
\end{eqnarray}
where (\ref{ok2}) is due to the fact that $W_1$ is a  function of $(A_{\mathcal{T}}, A_{\overline{\mathcal{T}}}^{[1]}, \mathcal{Q})$.  (\ref{ok3}) follows from the fact that $W_2$ is a  function of $(A_{\mathcal{T}}, A_{\overline{\mathcal{T}}}^{[2]}, \mathcal{Q})$. In (\ref{pk_1}), the third term is due to the fact that from $A_{\mathcal{T}}, A_{\overline{\mathcal{T}}}^{[3]}, \cdots, A_{\overline{\mathcal{T}}}^{[K]}$, we can decode $W_3, \cdots, W_K$.

Consider (\ref{pk_1}) for all subsets of $[1:N]$ that have exactly $T$ elements and average over all such subsets. We have
\begin{eqnarray}
&& KL - (K-2)L  + \frac{1}{\binom{N}{T}}\sum_{\mathcal{T}:|\mathcal{T}| = T}H(A_{\mathcal{T}}|W_1,W_2,\mathcal{Q}) \\
&\leq& N\mathcal{H}_T + \frac{1}{\binom{N}{T}}\sum_{\mathcal{T}:|\mathcal{T}| = T} \sum_{n \in \overline{\mathcal{T}}} H(A_n^{[2]} | A_{\mathcal{T}}, W_1,\mathcal{Q}) \\
&\leq& N\mathcal{H}_T + \frac{1}{\binom{N}{T}}\sum_{\mathcal{T}:|\mathcal{T}| = T} ~~\sum_{n \in \overline{\mathcal{T}}}~~\sum_{\mathcal{T'}\subset\mathcal{T}, |\mathcal{T'}|=T-1} \frac{H(A_n | A_{\mathcal{T}'}, W_1,\mathcal{Q}) }{T}\label{eq:s1}\\
&=& N\mathcal{H}_T  + \frac{1}{\binom{N}{T}}\sum_{\mathcal{T'}:|\mathcal{T'}| = T-1} ~~\sum_{n\notin\mathcal{T}'}(N-T)  \frac{H(A_n | A_{\mathcal{T}'}, W_1,\mathcal{Q}) }{T}\label{eq:s2}\\
&=& N\mathcal{H}_T  + \frac{1}{\binom{N}{T}}\sum_{\mathcal{T}:|\mathcal{T}| = T}\sum_{n\in\mathcal{T}} (N-T)  \frac{H(A_n | A_{\mathcal{T}/\{n\}}, W_1,\mathcal{Q}) }{T}\label{eq:s3}\\
&\leq&N\mathcal{H}_T + \frac{1}{\binom{N}{T}}\sum_{\mathcal{T}:|\mathcal{T}| = T}    (N-T)\frac{H( A_{\mathcal{T}}| W_1,\mathcal{Q})}{T}\label{eq:s4}\\
&\leq& N\mathcal{H}_T+  (\frac{N}{T} - 1) \frac{1}{\binom{N}{T}}\sum_{\mathcal{T}:|\mathcal{T}| = T} H(A_{\mathcal{T}}, A_{\overline{\mathcal{T}}}^{[1]} |  W_1,\mathcal{Q})   \\
&=&N\mathcal{H}_T+ (\frac{N}{T} - 1) \left( H(A_{[1:N]}^{[1]}, W_1  |  \mathcal{Q}) - H(W_1|\mathcal{Q}) \right)  \\
&=&N\mathcal{H}_T+ (\frac{N}{T} - 1) \Big( H(A_{[1:N]}^{[1]} |\mathcal{Q}) 
+ \underbrace{H(W_1| A_{[1:N]}^{[1]},\mathcal{Q})}_{= 0} - H(W_1) \Big)  \label{oky}\\
&\leq&N\mathcal{H}_T+ (\frac{N}{T}-1)  \left( N\mathcal{H}_T - L \right) \label{oksymm} \\
&=& \frac{N^2}{T} \mathcal{H}_T- (\frac{N}{T}-1) L  \\
\Rightarrow N\mathcal{H}_T &\geq& L \left( 1 + \frac{T}{N} \right)  +  \frac{T^2}{N^2} N \frac{1}{\binom{N}{T}}\sum_{\mathcal{T}:|\mathcal{T}| = T}\frac{H(A_{\mathcal{T}}|W_1,W_2,\mathcal{Q})}{T}  \label{okre}
\end{eqnarray}
where (\ref{eq:s1}) follows because dropping conditioning increases entropy. In (\ref{eq:s1}) we drop the superscript of $A_n^{[2]}$ and write it simply as $A_n$ because $\{A_n^{[2]},A_{\mathcal{T}'}\}$ contain responses from only $T$ databases, which must be independent of the desired message index because of the requirement of $T$-privacy. Equation (\ref{eq:s2}) follows from the observation that given $\mathcal{T}', n$, there are $(N-T)$ feasible choices of $\mathcal{T}$, i.e., those choices of $\mathcal{T}$ which contain $\mathcal{T}'$ and do not contain $n$. Thus, the same term $H(A_n|A_{\mathcal{T}'},W_1,Q)$ repeats in the overall summation a total of $(N-T)$ times. To see why (\ref{eq:s4})  holds, suppose we write $\mathcal{T}=\{n_1, n_2, \cdots, n_T\}$. By the chain rule of entropies, 
\begin{eqnarray}
H(A_{\mathcal{T}}|W_1,\mathcal{Q}) &=& \sum_{t=1}^TH(A_{n_t}|A_{n_{1}}, A_{n_{2}},\cdots, A_{n_{t-1}},W_1,\mathcal{Q})\\
&\geq& \sum_{t=1}^TH(A_{n_t}|A_{\mathcal{T}/{n_t}},W_1,\mathcal{Q})
\end{eqnarray}
In (\ref{oky}), the third term equals 0, because from $A_{[1:N]}^{[1]}$, we can decode $W_1$.

To proceed, we note that for the last term of (\ref{okre}), conditioning on $W_1, W_2$, the setting reduces to a PIR problem with $K-2$ messages and $N$ databases. Thus, (\ref{okre}) sets up an induction argument, which claims that for the $K$ messages setting,
\begin{eqnarray}
N\mathcal{H}_T  \geq L \left( 1 + \frac{T}{N} + \cdots + \frac{T^{K-1}}{N^{K-1}}  \right) \label{ind}
\end{eqnarray}
We have proved the basis cases of $K = 1$ and $K = 2$ in (\ref{eq:ind1}) and (\ref{re2}). Suppose now the bound (\ref{ind}) holds for $K-2$. Then plugging in (\ref{okre}), we have that the bound (\ref{ind}) holds for $K$. Since both the basis and the inductive step have been performed, by mathematical induction, we have proved that (\ref{ind}) holds for all $K$. The desired outer bound follows as
\begin{eqnarray}
R&=&\frac{L}{\sum_{n=1}^N H(A_n)}\leq \frac{L}{\sum_{n=1}^N H(A_n|\mathcal{Q})} \leq \frac{L}{N\mathcal{H}_T}  \leq \left(  1 + \frac{T}{N} + \cdots + \frac{T^{K-1}}{N^{K-1}} \right)^{-1}
\end{eqnarray} 
Thus, the proof of the outer bound  is complete.

\section{Proof of Theorem \ref{thm:robust}}\label{sec:robust}
Clearly the capacity of robust $T$-private PIR cannot be larger than the capacity of $T$-private PIR. Therefore, we only need to prove that the capacity of $T$-private PIR can be achieved in the robust PIR setting. To this end, we build upon the scheme presented in Section \ref{sec:ach} as follows. 

As before, let each message consist of $N^K$ symbols from a sufficiently large finite field $\mathbb{F}_q$. The messages  $W_1,\cdots,W_K\in\mathbb{F}_q^{N^K\times 1}$ are  represented as $N^K\times 1$ vectors over $\mathbb{F}_q$. Let $S_1, \cdots, S_K\in\mathbb{F}_q^{N^K\times N^K}$ represent random matrices chosen privately by the user, independently and uniformly from all $N^K\times N^K$  full-rank matrices over $\mathbb{F}_q$. Suppose $W_l$, $l\in[1:K]$, is the desired message. 

Consider any undesired message index $k\in [1:K]/\{l\}$, and all distinct $\Delta=2^{K-2}$  subsets of $[1:K]$ that contain $k$ and do not contain $l$. Assign distinct labels to each subset, e.g., $\mathcal{K}_1, \mathcal{K}_2, \cdots \mathcal{K}_{\Delta}$. 
For each $k\in [1:K]/\{l\}$, define the vector
{\setstretch{1.5}
\begin{eqnarray*}
\left[\begin{array}{l}x^{[k]}_{\mathcal{K}_1}\\
x^{[k]}_{\mathcal{K}_1\cup\{l\}}\\ \hdashline[2pt/2pt]
x^{[k]}_{\mathcal{K}_2}\\
x^{[k]}_{\mathcal{K}_2\cup\{l\}}\\ \hdashline[2pt/2pt]
\vdots\\ \hdashline[2pt/2pt]
x^{[k]}_{\mathcal{K}_\Delta}\\
x^{[k]}_{\mathcal{K}_\Delta\cup\{l\}}\\
\end{array}
\right]&=&\left[\begin{array}{llll}
\mbox{MDS}_{\frac{{\color{black}M}}{T}\alpha_1\times \alpha_1}&0&0&0\\ \hdashline[2pt/2pt]
0&\mbox{MDS}_{\frac{{\color{black}M}}{T}\alpha_2\times \alpha_2}&0&0\\ \hdashline[2pt/2pt]
0&\cdots&\ddots&0\\ \hdashline[2pt/2pt]
0&0&0&\mbox{MDS}_{\frac{{\color{black}M}}{T}\alpha_\Delta\times \alpha_\Delta}\\
\end{array}\right]S_k[(1:TN^{K-1}),:]W_k
\end{eqnarray*}
}
where $\alpha_i, i \in [1:\Delta]$ is defined as $N(N-T)^{|\mathcal{K}_i| - 1} T^{K - |\mathcal{K}_i|}$,  each $x^{[k]}_{\mathcal{K}_i}$ is a $\frac{M}{N} \alpha_i \times 1$ vector, and each $x^{[k]}_{\mathcal{K}_i\cup\{l\}}$ is a $\frac{M}{N}(\frac{N-T}{T})\alpha_i \times 1$ vector over $\mathbb{F}_q$. 

Now consider the desired message index $l$, and all distinct $\delta=2^{K-1}$ subsets of $[1:K]$ that contain $l$. Assign distinct labels to each subset, e.g., $\mathcal{L}_1, \mathcal{L}_2, \cdots \mathcal{L}_{\delta}$. Define the vector
{\setstretch{1.5}
\begin{eqnarray*}
\left[\begin{array}{l}x^{[l]}_{\mathcal{L}_1}\\
x^{[l]}_{\mathcal{L}_2}\\
\vdots\\
x^{[l]}_{\mathcal{L}_\delta}\\
\end{array}
\right]&=&{\color{black} \mbox{MDS}_{\frac{M}{N}N^K\times N^K}}S_lW_l
\end{eqnarray*}
where the length of $x_{\mathcal{L}_i}^{[l]}, i \in [1:\delta]$ is $M(N-T)^{|\mathcal{L}_i| - 1} T^{K - |\mathcal{L}_i|}$. 

For each non-empty subset $\mathcal{K}\subset[1:K]$ generate the query vector
\begin{eqnarray}
\sum_{k\in\mathcal{K}} {x}^{[k]}_\mathcal{K} 
\end{eqnarray}
Distribute the elements of the query vector evenly among the $M$ databases. This completes the construction of the queries.
}

Suppose the user collects answering strings from any $N$ databases. For each set $\mathcal{K}_i$, from $N$ databases, we download $\alpha_i$ symbols from $x^{[k]}_{\mathcal{K}_i}, k \neq l, i \in [1:\Delta]$, from which we can recover the interference $x^{[k]}_{\mathcal{K}_i \cup \{l\}}$, as they are generated by the generator matrix of a $(\frac{M}{T}\alpha_i,\alpha_i)$ MDS code. After subtracting out all the interference, we are left with $N^K$ desired symbols, from which we can recover the desired message, as the symbols are generated by the generator matrix of a $(\frac{M}{N}N^K, N^K)$ MDS code.
Therefore correctness is guaranteed.

Let us see why privacy holds. The queries for any $T$ colluding databases are comprised of $TN^{K-1}$ variables from each ${x}^{[k]}, k \in [1:K]$.  When $k = l$, the $TN^{K-1}$ desired symbols are generated by the generator matrix of a $(\frac{M}{N}N^K, N^K)$ MDS code such that these symbols have full rank. For each $k \neq l$, the $TN^{K-1} $variables from ${x}^{[k]}$ consist of $\alpha_i$ variables out of $\frac{M}{T}\alpha_i$  variables ${x}_{\mathcal{K}_i}^{[k]}, {x}_{\mathcal{K}_i\cup\{l\}}^{[k]}$, for each set $\mathcal{K}_i, i\in[1:\Delta]$. Note that these $\alpha_i$ variables are generated by the generator matrix of a $(\frac{M}{T}\alpha_{i}, \alpha_{i})$ MDS code, so that they have full rank.
Let the indices of the appeared variables be denoted by the  vectors $\mathcal{I}_{{x}^{[k]}}\in\mathbb{N}^{TN^{K-1}\times 1}, \forall k \in [1:K]$.
From Lemma \ref{lemma:inv}, we have
\begin{eqnarray}
{x}^{[k]}_{\mathcal{I}_{x^{[k]}}} &\sim& S_k[(1:TN^{K-1}),:]W_k
\end{eqnarray}
which  in turn implies that $S_k[(1:TN^{K-1}),:]$ are independent and identically distributed. Thus privacy is guaranteed.
Finally, the rate achieved is the same as that achieved in the setting without the robustness constraint. This completes the proof.

\section{Conclusion}\label{sec:conc}
We characterize the capacity of robust $T$-private PIR with arbitrary number of messages, arbitrary number of (responding) databases, and arbitrary privacy level. Let us conclude with a few  observations. First, while in this paper we adopt the zero error framework, we note that our converse extends in a straightforward manner to the $\epsilon$-error framework as well, where the probability of error is only required to approach zero as the message size approaches infinity. Therefore, for robust $T$-private PIR, the $\epsilon$-error capacity is the same as the zero error capacity. Second, recall that the capacity achieving scheme for PIR in our prior work \cite{Sun_Jafar_PIR} had a remarkable feature that if some of the messages were eliminated and the scheme projected onto a subset of messages, it remained capacity optimal for that subset of messages. The same phenomenon is observed for our achievable scheme for robust $T$-private PIR.
%A similar phenomenon can be observed for our achievable scheme for robust $T$-private PIR. %, provided that the number of messages that is eliminated is an even number. 
On the other hand, an important point of distinction of the  previous achievable scheme  in \cite{Sun_Jafar_PIR} from the achievable scheme in this paper is that the former directly uses each available side information symbol individually, whereas here we need MDS coded side information (uncoded side information symbols do not suffice). This is because of the $T$-privacy constraint which simultaneously creates multiple perspectives of external side-information depending upon which subset of databases decides to collude. 

Finally, we note that since we focus only on download cost,   upload cost is not optimized in this work. However,  even with $T$-privacy, significant optimizations of upload cost are possible through refinements of our achievable scheme. For example, the symbols may be grouped in a manner that randomizations are needed only within smaller groups, which may reduce the number of possible queries, and the size of the field of operations significantly. For example, consider the achievable scheme for $K=2, N=3, T=2$ that was presented in Section \ref{sec:k2n3t2}, where each message is comprised of $9$ symbols. We will operate over $\mathbb{F}_2$. Suppose we divide the $9$ bits into $3$ groups of $3$ bits each, and label the groups so that $A_1$ represents the first three bits of $W_1$, $A_2$ the next three and  $A_3$ represents the last three bits from $W_1$. Similarly, let $B_1, B_2, B_3$ represent three groups of three bits each from $W_2$.  Now, for any group of $3$ bits, say $X = (x_1, x_2, x_3)$, let $X(1), X(2), X(3)$ represent three randomly chosen linearly independent elements from the set $\{x_1, x_2, x_3, x_1+x_2, x_1+x_3, x_2+x_3,x_1+x_2+x_3\}$, i.e., selected  uniformly from  among the choices that do not sum to zero in $\mathbb{F}_2$. This essentially means that $X(1), X(2)$ may be freely chosen as any two distinct elements of the set and then $X(3)$ is chosen uniformly from the $4$ elements that are not $X(1), X(2)$ or $X(1)+X(2)$. The queries are  constructed as follows.

\begin{eqnarray*}
\begin{array}{|c|c|c|c|c|}\hline
\mbox{\tiny DB1}&\mbox{\tiny DB2}&\mbox{\tiny DB3}\\\hline
A_1(1),A_2(1)&A_2(2),A_3(2)&A_3(3),A_1(3)\\
B_1(1),B_2(1)&B_2(2),B_3(2)&B_3(1+2),B_1(1+2)\\
A_3(1)+B_3(1)&A_1(2)+B_1(2)&A_2(3)+B_2(1+2)\\
\hline
\end{array}
\end{eqnarray*}
where we use the notation $X(1+2)=X(1)+X(2)$ for brevity. Note that for the undesired symbols $B$, we used the $(2,3)$ MDS code  $(B(1), B(2))\longrightarrow (B(1),B(2),B(1+2))$ within each group. Due to the grouping of symbols the upload cost is significantly reduced. Moreover, because of the grouping we are able to operate over a smaller field. Whereas the original scheme presented in Section \ref{sec:k2n3t2} uses $(6,9)$ MDS codes which do not exist over $\mathbb{F}_2$, the refined example presented above uses only a $(2,3)$ MDS code which does exist over $\mathbb{F}_2$. As illustrated by this example, optimizations of upload costs as well as symbol size remain interesting avenues for  future work. 

\bibliographystyle{IEEEtran}
\bibliography{Thesis}
\end{document}